\documentclass{aa}
\input{psfig}
\begin{document}
%
%#########################################
%                title
%#########################################
%
%\thesaurus{03(
%            03.13.8;          % Methods: N-body simulations
%            03.13.2;          % Methods: data analysis
%            11.09.1 NGC 4449; % Galaxies: individual
%            11.09.1 DDO 125;  % Galaxies: individual
%            11.09.2;          % Galaxies: interactions
%            11.11.1)          % Galaxies: kinematics and dynamics
%               }

    \title{Multi-method-modeling of interacting galaxies.\\
           I. A unique scenario for NGC 4449?}
    \titlerunning{Multi-Method-Modeling of Interacting Galaxies. 
         I. A Unique Scenario for NGC 4449?}
 
\author{Christian Theis\inst{1} \& Sven Kohle\inst{2,}\inst{3} }
\authorrunning{Ch.\ Theis \& S. Kohle}
\institute{
   Institut f\"ur Theoretische Physik und Astrophysik der 
      Universit\"at Kiel, Olshausenstr.\ 40, 24098 Kiel, Germany,
      email: theis@astrophysik.uni-kiel.de \and
  Radioastronomisches Institut der Universit\"at Bonn, Auf dem H\"ugel 71,
      53121 Bonn, Germany \and
  Center for Medical Diagnostic Systems and Visualization MeVis GmbH at University of Bremen, Universit\"atsallee 29,
      28359 Bremen, Germany,
      email: sven@mevis.de}
\offprints{Ch.\ Theis}
\date{Received; Accepted}
%
%    #########################################
%                   abstract
%    #########################################
%
\abstract{
   NGC 4449 is an active star-forming dwarf ga\-la\-xy of Magellanic type.
  From radio observations, van Woerden et al.\ (\cite{vanwoerden75}) 
  found an extended 
  HI-halo around NGC 4449 which is at least a factor of 10 larger than the
  optical diameter $D_{\rm 25} \approx 5.6$ kpc. Recently,
  Hunter et al.\ (\cite{hunter98}) discerned details
  in the HI-halo: a disc-like feature around the center of NGC 4449 and
  a lopsided arm structure.\\
  We combined several N-body methods in order to investigate
  the interaction scenario between NGC 4449 and DDO 125, a close companion
  in projected space.
  In a first step fast restricted N-body models are used
  to confine a region in parameter space reproducing the main
  observational features. In a second step a genetic algorithm
  is applied for a uniqueness test of our preferred parameter set. 
  We show that our genetic algorithm reliably recovers
  orbital parameters, provided that the data are sufficiently
  accurate, i.e.\ all the key features are included.
  In the third step the results of the restricted N-body models
  are compared with self-consistent N-body simulations.
  In the case of NGC 4449, the applicability of the simple
  restricted N-body calculations is demonstrated. Additionally,
  it is shown that the HI gas can be modeled here by a purely stellar 
  dynamical approach.\\
   In a series of simulations, we demonstrate
  that the observed features of the extended HI disc can be explained 
  by a gravitational interaction
  between NGC 4449 and DDO 125. According to
  these calculations the closest approach between both galaxies
  happened $\sim 4-6 \cdot 10^8$ yr ago at a minimum distance of $\sim 25$ 
  kpc on a parabolic or slightly elliptic orbit. 
  In the case of an encounter scenario, 
  the dynamical mass of DDO 125 should not be smaller than 10\% of NGC 4449's
  mass. Before the encounter, the observed HI gas
  was arranged in a disc with a radius of
  35--40 kpc around the center of NGC 4449. It had the same 
  orientation as the central ellipsoidal HI structure. 
  The origin of this disc is still unclear, but it might have been caused by a 
  previous interaction.
}

\maketitle

%
%#########################################
%                text
%#########################################
%
%==============================================================
%                  Introduction
%==============================================================
\section{Introduction}

   During the last decades the classical picture of galaxies
being {\it Weltinseln} (e.g.\ Kant \cite{kant55};
v.\ Humboldt \cite{humboldt50}),
i.e.\ island universes, has completely changed. 
In the 1920s, galaxies were thought to be
in general {\it islands}, i.e.\ isolated stellar systems, with just a
few obvious exceptions like the M51 system. 
Systems not fitting the Hubble classification have been neglected  
-- by definition -- as peculiar. Interest in these systems has increased 
strongly since the new catalogues
by Vorontsov-Velyaminov (\cite{vorontsov59}) and Arp (\cite{arp66}) 
became available. They demonstrated that even peculiar objects have 
common features like bridges, tails, or rings.
Better observational instruments allowing for deeper images 
and new wavelength ranges have revealed more complex structures.
Galaxies formerly classified as non-interacting may have had
some gravitational interaction in the past which is still reflected
in e.g.\ warps, flares, or thick discs. Especially useful are 
HI observations which can cover a much larger radial extension than
the optical images. Therefore, HI is a very good tracer for tidal interactions.

  The theoretical understanding of interacting galaxies suffered
for a long time from the lack of computational power allowing for
a numerical solution of the gravitational N-body problem. After
a remarkable treatment by Holmberg (\cite{holmberg41}) 
who built an analogue computer
(consisting of light bulbs and photo cells) to determine the gravitational
force, it took 20 years until N-body simulations were performed on a general
purpose computer (Pfleiderer \& Siedentopf \cite{pfleiderer61}): 
The basic idea
of these {\it restricted N-body} simulations is the assumption that the
potential of interacting galaxies can be adequately modeled by
two particles which represent the galactic masses and move under their
mutual gravitation, i.e.\ on Keplerian orbits. With these
assumptions all the other particles are just test particles, and the 
complete N-body problem is reduced to $N$ single body problems for a
time-dependent potential. In a remarkable series of simulations
Toomre \& Toomre (\cite{toomre72}) applied this technique to determine
the parameters of some well-studied interacting
systems like Arp 295, M 51 + NGC 5195, or NGC 4038/39. 

New N-body techniques have been developed which increased the accessible
particle number by many orders of magnitude. E.g.\
the TREE-method (Barnes \& Hut \cite{barnes86}; 
Hernquist \cite{hernquist87}, \cite{hernquist90})
in which the
organization of the force calculation -- the most time-consuming part
of N-body calculations -- adapts to clumpy mass distributions for a moderate
computational price ($\sim O( N\log N )$) compared to direct simulations
($\sim O(N^2)$). By this, Barnes (\cite{barnes88}) 
was able to simulate encounters
of disc galaxies including all dynamical components, i.e.\ the disc,
the bulge, and the halo as N-body systems.
Compared to faster grid-based methods (e.g.\ Sellwood \cite{sellwood80})
or expansion methods (e.g.\ Hernquist \& Ostriker \cite{hernquist92}) 
direct N-body 
simulations (or semi-direct methods like TREEs) are more flexible 
with respect to strongly varying geometries and scalelengths. 
An alternative to these techniques are special-purpose computers 
(such as the machines of the GRAPE project 
(Sugimoto et al.\ \cite{sugimoto90})). They implemented Newton's law of gravity
(modified for gravitational softening) in the hardware which allows for
a very fast direct determination of the gravitational forces (however, 
there still exists the $N^2$ bottleneck). E.g.\ for simulations with 
$N=10^5$ particles a GRAPE3af (with 8 GRAPE processors) is 
competitive with a TREE-code running on a CRAY T90. 

  Two main methods (based on particles) to treat a dissipative, 
star-forming gas have been applied:
The {\it smoothed particle hydrodynamics} scheme (SPH)
solves the gasdynamical equations, by this emphasizing the 
diffuse nature of the interstellar medium (ISM)
(e.g.\ Hernquist \& Katz \cite{hernquist89}). 
An alternative approach focuses
on the clumpiness of the ISM treated as
{\it sticky particles}: Without physical collisions or close encounters 
of clouds they move 
in ballistic orbits like stars. However, in the case of a collision
the clouds might merge or lose kinetic energy depending on
the adopted microphysics (e.g.\ Casoli \& Combes \cite{casoli82};
Palou\v s et al.\ \cite{palous93}; Theis \& Hensler \cite{theis93}).

Relatively few papers exist modeling special
objects. This deficiency is the result of two factors: First
high resolution data in configuration and velocity space are required,
covering a large 
fraction of the space between the interacting galaxies. In principle, 
HI would be suitable, however, there are just
a few observational sites which give data of sufficient quality. 

 The second problem is the large parameter space for a galactic interaction
resulting in two connected difficulties: finding a good fit and
determining its uniqueness (or other acceptable parameter sets).
Observationally, only three kinematical quantities -- the projected position 
on the sky and the line-of-sight velocity -- can be measured. Another parameter,
the galactic mass, depends on the availability of velocity data, the 
determination of the distance, and the reliability of the conversion from
velocity to masses. Neglecting the center-of-mass data of the interacting
system these 14 parameters reduce to 7 parameters containing the relative
positions and velocities. These 7 values just fix the orbit in the case
of a two-body interaction. Moreover, one has to specify the
parameters that characterize both stellar systems, e.g.\ characteristic
scales, orientation, or rotation. The final result is
a high-dimensional parameter space which is in general too large for
a standard search method. For instance, the interaction
of a galactic disc with a point-mass galaxy is described at least 
by 7 parameters. A regular grid with a poor coverage of 10
grid points per dimension demands $10^7$ models or 3400 years GRAPE
simulation time (assuming 3 CPU-hours for a single simulation) or
still about a year with a restricted N-body program (assuming 3 CPU-seconds
per simulation). 

An alternative approach to investigate large
parameter spaces is a genetic algorithm (Holland \cite{holland75};
Goldberg \cite{goldberg89}). This applies an evolutionary mechanism 
on a population (i.e.\ a set of parameters), from which members 
are selected for
parenthood according to their {\it fitness}. Fitness is determined
by the ability of the parameters to match the observations,
whereas the reproduction within the genetic algorithm 
is done by {\it cross-over} of the parental
parameters and applying a {\it mutation} operator. 
Although genetic algorithms (GA)
have been used in many branches of science, there are just a few
applications in astrophysics, e.g.\ for fitting rotation curves or
analysis of Doppler velocities in $\delta$ Scuti stars (for a review see 
Charbonneau \cite{charbonneau95}). Recently, Wahde (\cite{wahde98}) 
demonstrated the ability
of genetic algorithms to recover orbital parameters for artificial
data generated by N-body simulations of strongly interacting galaxies. 
He stressed that in these systems the positional information can be
already sufficient to determine the orbital elements.

   In this paper we combine different N-body techniques
in order to create a model for the dynamics of interacting
galaxies. In a first step, restricted N-body simulations are performed
to constrain the model parameters: If a sufficiently accurate data set is
available, the genetic algorithm can be used to find the orbital
parameters and to check the uniqueness of the result. Additionally,
the sensitivity of the solution can be checked by an extended parameter 
study. Alternatively, in the case of insufficient data, 
one can use the restricted N-body models to obtain
a first guess of the parameters, creating an artificial 
intensity map and then apply the genetic algorithm to check its uniqueness.
In the second step, self-consistent models are necessary in order
to tune the parameters and to check the applicability of the restricted 
N-body method. These checks must address two questions: Is the 
neglection of self-gravity acceptable? Does gas dynamics alter 
the results significantly? Technically this means that self-consistent
simulations with and without gas have to be performed. 

  The numerical models of this paper are performed on NGC 4449
which is a prototype actively star-forming galaxy similar
in size and mass to the Large Magellanic Cloud. NGC 4449
is surrounded by an extended HI halo which shows some distortion
(van Woerden et al.\ \cite{vanwoerden75}; Bajaja et al.\ \cite{bajaja94}). 
Hunter et al.\ (\cite{hunter98})
published high resolution HI images showing a 
large scale lopsided structure. Though there seems to be
a close dwarf companion, DDO 125, it is not clear whether the observed
streamers are tidal tails caused by an interaction with DDO 125
(Hunter et al. \cite{hunter98}). For example, tidal distortions 
are missing in the optical 
image of both galaxies and no clear bridge has been detected. 
Theis \& Kohle (\cite{theis98}) 
demonstrated qualitatively by N-body simulations that the main
HI features could stem from a face-on encounter provided the mass
of DDO 125 exceeds 10\% of the mass of NGC 4449. An alternative 
scenario assumes that the HI structure is an ongoing infall
of gas onto NGC 4449, possibly caused by the encounter
with DDO 125 followed by a compression and cooling of the gas 
(Silk et al.\ \cite{silk87}). However, it is puzzling
that the observed large-scale, lopsided and regular structure 
in NGC 4449 can be maintained over timescales such as the 
orbital period which is longer than 1 Gyr for the outer streamer.
The high internal velocity dispersion of 10 km\,s$^{-1}$ would destroy
the streamers on a timescale of $2 \cdot 10^8$ yr (Hunter et al.\
\cite{hunter98}).
In this paper we investigate this interaction scenario, addressing the 
question of whether the observed structures can actually be formed by tidal
interaction and which constraints on the mass distribution
or the orbits of both galaxies can be derived.

Sect.\ \ref{observations} summarizes the observational data of NGC 4449 
and DDO 125 relevant for the models of this paper. The different
modeling techniques and the numerical results are described
in Sect.\ \ref{numerics} (the restricted N-body
method is introduced (Sect.\ \ref{restricted}) and a set of
first results is shown in Sect.\ \ref{firstguess} and \ref{parameterstudy}).
The genetic algorithm approach is briefly 
introduced in Sect.\ \ref{geneticalgorithm} and explained in more detail
in the Appendix. Its results are shown in Sect.\ \ref{unique}. 
Finally, the restricted N-body simulations are compared with 
self-consistent models with (Sect.\ \ref{stickyparticles}) and 
without gas (Sect.\ \ref{scnogas}). The results of these different 
approaches are summarized and discussed in Sect.\ \ref{discussion}.
%
%==============================================================
%                  The case of NGC 4449
%==============================================================
\section{The case of NGC 4449 and DDO 125}
\label{observations}
%
%==============================================================
%                  Observations
%==============================================================
\subsection{NGC 4449}

   When optical images of NGC 4449 and the Large Magellanic
Cloud are compared, many similarities like the presence of 
a bar, the size or the total luminosity are found. 
However, unlike the LMC
NGC 4449 is a candidate for studies of the properties of a 
Magellanic-type irregular unaffected by the tidal field of a 
large galaxy. Its distance of
2.9 -- 5.6 Mpc\footnote{Most of the distance estimates for NGC 4449 differ 
because of the assumed value $H_0$ for the Hubble constant. 
Independently of $H_0$, Karachentsev \& Drozdovsky (\cite{karachentsev98}) 
derived from the photometry of the 
brightest blue stars a distance of 2.9 Mpc. However, the error of this
measurement is unclear. In accordance with Hunter 
et al.\ (\cite{hunter98}) a distance of 3.9 Mpc will be assumed in this paper.
Thus, $1^\prime$ corresponds to 1.1 kpc.} 
allows for detailed investigations, performed in the
visual (e.g.\ Crillon \& Monnet \cite{crillon69}; 
Hunter \& Gallagher \cite{hunter97}), in the radio band 
(e.g.\ van Woerden et al.\ \cite{vanwoerden75}; Klein et al.\ \cite{klein96}),
in the HI emission line (e.g.\ Bajaja et al.\ \cite{bajaja94}; 
Hunter \& Gallagher \cite{hunter97}; Hunter et al.\ \cite{hunter98}), 
in the CO emission line (e.g.\ Sasaki et al.\ \cite{sasaki90}; 
Hunter \& Thronson \cite{hunter96}; Kohle et al.\ \cite{kohle98}), 
in the infrared (e.g.\ Hunter et al.\ \cite{hunter86}), 
in the FUV (e.g.\ Hill et al.\ \cite{hill94}) and in
X-rays (e.g.\ della Ceca et al.\ \cite{dellaceca97}; 
Bomans \& Chu \cite{bomans97}). Inside the Holmberg diameter of 11 kpc,
NGC 4449 shows ongoing strong star formation along a bar-like 
structure (Hill et al.\ \cite{hill94}). This activity seems to induce a 
variety of morphological features like filaments, arcs, or loops which have
a size of several kpc (Hunter \& Gallagher \cite{hunter97}).

%---------------------------
%  fig1
%---------------------------
\begin{figure}[t]
      \vspace*{8.5cm}
%     \centerline{\psfig{figure=fig1.ps,height=8.5cm,width=8.5cm}}
     \caption[ ]{\it
       Contour map of the integrated HI-intensity of NGC 4449. NGC 4449 is
       at the center of the image and DDO 125 is located at the
       almost circular contours in the South. Details are
       given in Hunter et al.\ (\cite{hunter98}). This image
       was kindly provided by Deidre Hunter.}
     \label{hunterobservation}
\end{figure}

Early radio observations
of van Woerden et al.\ (\cite{vanwoerden75}) exhibited a large HI structure
exceeding the Holmberg diameter of 11 kpc by a factor of 7.5.
Recent observations by Bajaja et al.\ (\cite{bajaja94}) 
with the Effelsberg telescope 
and by Hunter et al.\ (\cite{hunter98}) with
the VLA revealed a complex structure of several components (Fig.\ 
\ref{hunterobservation}):
An elongated ellipse of HI gas centered on NGC 4449 has a total mass
of $\sim 1.1 \cdot 10^9 \, {\rm M}_\odot$. This gas rotates rigidly inside
11 kpc, reaching a level of 97 km\,s$^{-1}$ in the deprojected velocity.
Outside 11 kpc, the rotation curve is flat. Bajaja et al.\ (\cite{bajaja94})
derived a dynamical mass of $ 7.0 \cdot 10^{10} \, {\rm M}_\odot$ inside 32 kpc and a HI 
gas mass fraction of 3.3\%. Within 11 kpc the dynamical mass
is $2.3 \cdot 10^{10} \, {\rm M}_\odot$ and the mass inside the optical radius 
of about 5.6 kpc is $3 \cdot 10^{9} \, {\rm M}_\odot$. Applying single dish data for 
a $21^\prime$ beamwidth, Hunter et al.\ (\cite{hunter98}) 
derived about $10^{10} \, {\rm M}_\odot$. The HI gas inside the optical part of 
NGC 4449 counterrotates with respect to the outer regions.
For the central ellipse, Hunter et al.\ (\cite{hunter98}) determined a
position angle of $230^\circ \pm 17^\circ$ and an inclination of
$60^\circ \pm 5^\circ$.

South of the galactic center, a streamlike 
structure of 25 kpc emanates which abruptly splits into two parts:
A small spur (5 kpc) which points towards DDO 125, a close irregular
galaxy, and a long extended stream. 
The latter first points 24 kpc to the
north (in the following: {\it vertical streamer}), 
then 49 kpc north-east and, finally, 27 kpc to the south-east
(together called the {\it northeastern streamer}),
covering 180$^\circ$ around NGC 4449. The long
straight sections are notable, as are the rather abrupt changes of direction. 
The total mass of the streamers measured by the VLA observations
is about $1 \cdot 10^{9} \, {\rm M}_\odot$. Hunter et al.\ (\cite{hunter98}) point
out that compared to the Effelsberg data there is an additional 
diffuse HI mass not detected in the VLA observations which amounts to
two-thirds of the HI in the extended structures.

\subsection{DDO 125}

   DDO 125 is located in the south-east of NGC 4449 at a projected
distance of 41 kpc. The line-of-sight velocity difference between DDO 125
and NGC 4449 is only about 10 km\,s$^{-1}$. Tully et al.\ (\cite{tully78}) found
rigid rotation inside the optical radius of DDO 125 and an inclination of the
adopted disc of $50^\circ$. From this they derived a total mass of 
$5 \cdot 10^{8} \, {\rm M}_\odot$ (corrected for a distance of 3.9 Mpc) which is
in agreement with single dish observations ($21^\prime$ beamwidth) 
of Fisher \& Tully (\cite{fisher81}). However, Ebneter et al.\ 
(\cite{ebneter87}) reported VLA-observations which show a linear
increase of the rotation curve out to the largest radii where HI
has been detected, i.e.\ out to twice the optical radius. This already
allows for an estimate of a lower mass limit for DDO 125 
which is a factor 8 larger than
the value of Tully et al.\ (\cite{tully78}), i.e.\ $4 \cdot 10^9 \, {\rm M}_\odot$. Since
neither flattening of the rotation curve nor any decline has been observed,
this value is only a lower limit. If one compares that mass with the 
{\it total} mass of NGC 4449 derived by Bajaja et al.\ (\cite{bajaja94}) 
one gets a mass
ratio of about $q \approx 5.7\%$ which is, however, just a lower limit. 
Comparison of the masses inside the range of rigid rotation gives 
a higher mass ratio of $q \approx 18\%$.

%
%==============================================================
%                  Numerical Modeling
%==============================================================
\section{Numerical Modeling}
\label{numerics}

For the simulations described in this paper three different N-body 
approaches have been applied: The method of restricted N-body simulations
(Pfleiderer \& Siedentopf \cite{pfleiderer61}; Toomre \& Toomre
\cite{toomre72}) is described
in Sect.\ \ref{restricted}. The self-gravitating models are 
split into calculations without gas (Sect.\ \ref{scnogas}) and simulations
which include gas by means of sticky particles (Sect.\ \ref{stickyparticles}).

  The advantage of the first method is twofold: the reduction of the full 
N-body problem to $N$ single body problems and the use of particles as 
test particles which gives high spatial resolution at the
places of interest for almost no computational cost. As a direct consequence,
restricted N-body calculations can be performed with fewer particles
than self-consistent simulations which strongly suffer from
two-body relaxation in case of small particle numbers. Furthermore,
the CPU-time scales linearily with particle number which is
superior to almost all N-body codes. E.g.\ a restricted N-body simulation
sufficiently reproduces tidal features with only 1000 particles,
whereas a direct self-consistent calculation
needs an (optimistic) minimum of 10\,000 particles to reproduce a 
stable galactic disc on the interaction timescale. Comparing the CPU-times 
of such a 
self-consistent simulation performed on a GRAPE3 board with the values of a 
restricted N-body simulation we get a time saving of at least a factor of
1000.\footnote{The exact acceleration depends strongly on the details
of the comparison, e.g.\ the requirements on disc stability and therefore
the number of particles for the self-consistent model, the detailed
orbits, the method of the self-consistent calculation etc. An acceleration
of a factor 1000 is only a lower limit.}
Hence, the restricted N-body calculation
is at the moment the only method which allows about $10^4$
simulations in a couple of CPU-hours. Such a fast computation is a 
prerequisite for all systematic searches in parameter space such as the 
genetic algorithm described in Sect.\ \ref{geneticalgorithm}.

However, the simplified treatment of the galactic potential in 
restricted N-body simulations prohibits
any feedback of the tidal interaction on the orbits of both
galaxies, e.g.\ no transfer of orbital angular momentum into galactic spin
is possible (i.e. no merging). To overcome this 
problem, the results of restricted N-body models and genetic algorithms
must be compared with detailed self-consistent simulations. In a first
step the stellar dynamical applicability of the restricted models is
checked by a comparison with self-gravitating systems calculated with
a GRAPE3af special purpose computer (Sugimoto et al.\ \cite{sugimoto90}; see
Sect.\ \ref{scnogas}). 

   So far the dynamics of the system have been investigated in terms of
stellar dynamics, though in many cases (such as the
case of NGC 4449) HI {\it gas} is observed. In general, the effects of 
neglecting gas dynamics are not clear. E.g.\ a purely stellar dynamical 
ansatz was successfully applied for the system M 81 and NGC 3077 
(Thomasson \& Donner \cite{thomasson93}), whereas in other systems 
the dynamics of the gas can deviate strongly
from the behaviour of the stars (e.g.\ Noguchi \cite{noguchi88};
Barnes \& Hernquist \cite{barnes96}). 
Therefore, one has to compare stellar and gas dynamical results
which is done in this paper by a sticky particle method (Sect.\ 
\ref{stickyparticles}).\\

The units are chosen as 1 kpc, $7 \cdot 10^{10} \, {\rm M}_\odot$ 
(the dynamical mass of NGC 4449) and G=1. This results
in a time unit of 1.78 Myr and a velocity of 549 km\,s$^{-1}$.

%
%==============================================================
%                  Restricted N-body model
%==============================================================
\subsection{Restricted N-body models}
\label{restricted}

  Restricted N-body simulations aim to
reduce the N-body problem to $N$ 1-body problems by assuming that
the gravitational potential is given by a simple relation. E.g.\
two- or a few-body problems have known \mbox{(semi-)}analytical 
solutions or can be solved by fast standard methods. Here
we assume that the gravitational forces on each particle are given
by the superposition of the forces exerted by two point-like objects
(the galaxies) moving on Keplerian orbits. Thus, for the orbits one has to
specify 14 parameters which reduce to 7 in the center-of-mass frame
(relative coordinates/velocities and the mass ratio): 
The orbital plane is fixed by the inclination angle and the argument of 
the ascending node. The orbit itself is characterized by its
eccentricity and the location
of the pericenter, i.e.\ the pericentric distance and the argument
of the pericenter. Finally, mass and initial location of the
{\it reduced} particle (or the time of pericentric passage) have 
to be specified to fix the phase. For a comparison with the observations,
the phase of the observation also must be supplied. 
Three of the seven orbital parameters can be fixed
by the observed location of both galaxies on the plane of sky 
($x$-$y$ plane) and the line-of-sight velocity. The masses (and thus
the mass ratio) can be estimated by the velocity profiles of both 
galaxies, provided the distance to them is known. 

In this paper, the time of pericentric passage is set to zero. As input,
the orbital eccentricity and inclination and the minimum distance are
given. Furthermore, the directly
observable parameters, i.e.\ the masses of the galaxies, the projection of
the relative positions of both galaxies onto the plane of sky, and their
line-of-sight velocity difference, are supplied. They are used to
determine the argument of the pericenter, the phase (or time since
pericentric passage) of the observed system and the argument of
the ascending node of the orbital plane. Additionally, the initial distance
or phase has to be specified: In the case of parabolic and hyperbolic
encounters, the distance was set to 100 kpc, which guarantees a sufficiently
low tidal influence at the start of the simulation. In the case of an elliptical
encounter, the situation is more difficult, because the apocentric distance 
might be small, especially if the eccentricity is low. Thus, the system 
might be tidally affected all the time or by a series of 
repeated encounters which makes the applicability of the restricted N-body 
method doubtful. For these elliptical orbits, the initial azimuthal angle of 
the orbit was set to 120$^\circ$ (0$^\circ$ corresponds to pericenter).

  The test particles are arranged in a flat disc moving on circular
orbits. The disc itself is characterized by its orientation, i.e.\
its inclination and position angle, the scalelength and the outer edge.
Additionally, an inner edge can be specified in order to avoid a waste
of computational time by integrating tidally unaffected orbits well
inside the tidal radius of the galaxy.
  In most of the restricted N-body simulations the test particles are 
distributed in 30 to 60 equidistant rings of 125 particles, i.e.\ a total
of $N=3750$ to $7500$ particles. The typical spacing
of the rings is 0.7 to 1.5 kpc depending on the outer edge of the disc.
The time integration of the equations of motion 
is performed by a Cash-Karp Runge-Kutta integrator of fifth 
order with adaptive timestep control (Press et al.\ \cite{press92}). In order
to prevent a vanishingly small timestep by an accidental infall of a test
particle to the galactic center, the gravitational force is
softened on a length scale of 1 kpc. In order to speed-up
the calculation the galactic positions are tabulated by cubic splines
and the integrator uses these look-up tables. On a SPARC10 with a 
150 MHz CPU a typical 7500-particle encounter takes 23 CPU seconds,
which can be reduced by a factor of 2-3 by applying an inner edge
of 10 kpc. 

%
%===================================================================
%                  Results of the restricted N-body calculations
%===================================================================
%
%
%==============================================================
%                  A first guess
%==============================================================
%
\subsection{A first guess}
\label{firstguess}

%---------------------------
%  fig2
%---------------------------
\begin{figure}
    \vspace*{11.5cm}
   \caption[ ]{\it Temporal evolution of the particle positions 
     around NGC 4449 projected onto the plane-of-sky for different times. 
     The perturber DDO 125 is shown by the star. Closest approach is at 
     $t=0$. The final diagram corresponds to a projected distance of 
     41 kpc. The mass ratio of both galaxies is set to $q=0.2$. The original
     HI disc has an inclination of $i=60^\circ$, a position angle
     $\alpha = 230^\circ$ and a radial extent of $R_{\rm max}=40$ kpc. 
     The particles inside a radius of 10 kpc are omitted. The contour 
     lines in the lower right diagram show the normalized surface density 
     distribution corresponding to the particle distribution in the lower 
     left diagram. The intensities are calculated on a 50$\times$50 grid and
     smoothed over 3 grid cells in each dimension. One time unit
     corresponds to 1.78 Myr.
      }
     \label{resevol}
\end{figure}

   In a series of simulations the orbital parameters
(minimum distance $d_{\rm min}$, the eccentricity $\epsilon$, the mass
ratio $q$ and the orientation of the orbital plane) as well as the 
disc parameters (size $R_{\rm max}$ of the disc, 
its position angle $\alpha$ and inclination $i$) 
are varied (see Table \ref{modeltable}). 

%The temporal evolution of the primary galaxy
The evolution of the primary galaxy (NGC 4449) in the reference model A 
is displayed in Fig.\ \ref{resevol}. At the beginning, both galaxies
have a distance of 100 kpc corresponding to 995 Myr before closest
approach. The particles inside 10 kpc are omitted for an easier 
identification of the evolving structure. This does not
alter the results, since the related particles are only 
weakly affected by the interaction, as the persistence of the
elliptical shape of the central hole demonstrates. At the
moment of closest approach ($t=0$), only the outer region of
NGC 4449's disc reacts on the intruder. 71 Myr later
(i.e.\ system time 40),
a dense outer rim,  starting at the projected position of
DDO 125 and pointing to the north, begins to form. Secondly,
a weak bridge-like feature seems to connect both galaxies.
Both features are more pronounced after 178 Myr.
At that time an additional tidal feature begins to evolve
north of the center. When DDO 125 has a projected distance of 41 kpc 
(362 Myr after closest approach), three features are discernible:
the remnant of the former bridge starts south of the center of
NGC 4449 pointing to the south-west. The vertical feature starts
at the end of the first structure and points to the north.
The third structure is a long tidal arm which starts west of NGC 4449
pointing to north-east before turning back to the south-east.

A comparison of the intensity map of model A (lower right diagram in
Fig.\ \ref{resevol}) with the observations (Fig.\ \ref{hunterobservation}) 
demonstrates that the main observational features
can be reproduced qualitatively by the reference model A, i.e.\ a
parabolic orbit, a mass ratio $q=0.2$ and a minimum distance 
$d_{\rm min}=25$ kpc.
The parameters for the disc orientation are chosen in agreement 
with Hunter et al.'s suggestion of $i=60^\circ$, $\alpha=230^\circ$,
whereas the disc radius was set to $R_{\rm max}=40$ kpc.
Although this numerical model is not a detailed fit of the data, 
the characteristic sizes and locations of the streamers as well 
as the unaffected disc can be found. The material located between the 
streamer and the disc, which is not seen in Hunter's data, 
corresponds to the extended HI mass detected in the single dish 
observations by van Woerden et al.\ (\cite{vanwoerden75}). 
In the following sections 
we will discuss the influence of the individual parameters. The global
uniqueness of the reference model is investigated in
Sect.\ \ref{unique}.

%
%==============================================================
%                  Parameter study
%==============================================================
\subsection{Parameter study}
\label{parameterstudy}

  In this section the influence of the basic parameters, 
i.e.\ the orientation of the orbital plane, the orbital eccentricity
and the minimum distance, the mass ratio of both galaxies, and the orientation
of the HI disc and its extent are discussed (see Table \ref{modeltable}).
In the following simulations, all parameters except the one of interest
are kept constant with respect to model A. Thus, the vicinity of the
reference model in parameter space is studied.\\

\begin{table*}
  \label{modeltable}
  {\bf Model parameters of the restricted N-body simulations}

  \vspace*{0.1cm}

  \begin{tabular}{|l || c|c|c|c || c|c|c || l|}
    \hline
    model & $\epsilon$ & $q$  & $d_{\rm min}$ & $i$ & $i_d$ & $\alpha_d$ & $R_{\rm max}$ & comment \\ \hline\hline
     A    &  1.0       & 0.2  &  25           & 40  &   60  & 230        &  40  & reference model\\ \hline\hline

     B1   &  0.5       & 0.2  &  25           & 40  &   60  & 230        &  40  & eccentricity \\ \hline
     B2   &  0.8       & 0.2  &  25           & 40  &   60  & 230        &  40  & \\ \hline
     B3   &  1.5       & 0.2  &  25           & 40  &   60  & 230        &  40  & \\ \hline
     B4   &  5.0       & 0.2  &  25           & 40  &   60  & 230        &  40  & \\ \hline\hline

     C1   &  1.0       & 0.01 &  25           & 40  &   60  & 230        &  40  & mass ratio \\ \hline
     C2   &  1.0       & 0.05 &  25           & 40  &   60  & 230        &  40  & \\ \hline
     C3   &  1.0       & 0.1  &  25           & 40  &   60  & 230        &  40  & \\ \hline
     C4   &  1.0       & 0.3  &  25           & 40  &   60  & 230        &  40  & \\ \hline
     C5   &  1.0       & 0.4  &  25           & 40  &   60  & 230        &  40  & \\ \hline\hline

     D1   &  1.0       & 0.2  &  15           & 40  &   60  & 230        &  40  & minimum distance \\ \hline
     D2   &  1.0       & 0.2  &  20           & 40  &   60  & 230        &  40  & \\ \hline
     D3   &  1.0       & 0.2  &  30           & 40  &   60  & 230        &  40  & \\ \hline
     D4   &  1.0       & 0.2  &  35           & 40  &   60  & 230        &  40  & \\ \hline
     D5   &  1.0       & 0.2  &  40           & 40  &   60  & 230        &  40  & \\ \hline\hline

     E1   &  1.0       & 0.2  &  25           & 10  &   60  & 230        &  40  & orbital inclination \\ \hline
     E2   &  1.0       & 0.2  &  25           & 20  &   60  & 230        &  40  & \\ \hline
     E3   &  1.0       & 0.2  &  25           & 30  &   60  & 230        &  40  & \\ \hline
     E4   &  1.0       & 0.2  &  25           & 50  &   60  & 230        &  40  & \\ \hline
     E5   &  1.0       & 0.2  &  25           & 70  &   60  & 230        &  40  & \\ \hline\hline

     F1   &  1.0       & 0.2  &  25           & 40  &   40  & 230        &  40  & disc inclination \\ \hline
     F2   &  1.0       & 0.2  &  25           & 40  &   50  & 230        &  40  & \\ \hline
     F3   &  1.0       & 0.2  &  25           & 40  &   70  & 230        &  40  & \\ \hline\hline

     G1   &  1.0       & 0.2  &  25           & 40  &   60  & 210        &  40  & pos.\ angle of disc \\ \hline
     G2   &  1.0       & 0.2  &  25           & 40  &   60  & 250        &  40  & \\ \hline\hline

     H1   &  1.0       & 0.2  &  25           & 40  &   60  & 230        &  25  & outer edge of disc \\ \hline
     H2   &  1.0       & 0.2  &  25           & 40  &   60  & 230        &  30  & \\ \hline
     H3   &  1.0       & 0.2  &  25           & 40  &   60  & 230        &  35  & \\ \hline
     H4   &  1.0       & 0.2  &  25           & 40  &   60  & 230        &  50  & \\ \hline
  \end{tabular}
  \caption[]{\it Parameters of the different restricted N-body models: 
     model name \mbox{(col.\ 1)}, orbital eccentricity
   $\epsilon$ (col.\ 2), mass ratio $q$ of both galaxies 
    (col.\ 3), minimum separation $d_{\rm min}$ of galactic centers
    (in kpc, col.\ 4), inclination $i$ of orbital plane 
    (in degrees, col.\ 5), orientation of NGC 4449 disc (in degrees):
    inclination $i_d$ (col.\ 6) and position angle $\alpha_d$ (col.\ 7), 
    maximum radius $R_{\rm max}$ of gaseous disc \mbox{(in kpc, col.\ 8)}.}
\end{table*}

%---------------------------
%  fig3
%---------------------------
\begin{figure}
    \vspace*{12.8cm}
   \caption[ ]{\it Dependence on the eccentricity.
     Projection of the particle positions onto the plane-of-sky at the 
     moment of a projected distance of 41 kpc. The eccentricities of the models
     are 0.5 (upper left, model B1), 0.8 (upper right, B2),
     1.0 (A, reference model), 1.5 (B3) and 5.0 (B4).
      }
     \label{resecc}
\end{figure}

\noindent
{\bf Eccentricity.} The eccentricity mainly influences the interaction
timescale which is defined here as the time between maximum approach and
a projected distance of 41 kpc, i.e.\ the moment of observation. 
An eccentricity of 5.0 (corresponding to a hyperbolic encounter)
decreases the interaction time by
a factor 2.6 resulting in less structural changes of the primary galaxy
(Fig.\ \ref{resecc}): Only the remnant of the bridge feature can be 
discerned, whereas the other two streamers found in the reference model
are missing completely. With decreasing eccentricity, the north-eastern
streamer becomes pronounced before the vertical structure arises. The latter
is weakly expressed for $\epsilon=1.5$. Reducing $\epsilon$ further
to bound orbits makes the features even more prominent. In the case
of $\epsilon=0.5$ the north-eastern and the vertical streamer form
a single structure clearly detached from the center.
However, due to the strength of the interaction, a second broad tidal arm
is ejected from NGC 4449, leaving a low-density region south to the 
northern tip of the north-eastern streamer.

 In order to compare the $\epsilon=0.5$ model with observations,
we reran model B1, but with $N=80\,000$, in order to derive
intensity and velocity maps. This model reproduces several 
observed features very well (Fig.\ \ref{reseccdetail}):
First, the intensity map shows the three observed abrupt changes: 
at the lower tip of the vertical streamer, at the transition from the
vertical streamer to the north-eastern streamer and at the northern tip
of the north-eastern streamer. Second, the spatial extent of the streamers
is in agreement with the observations. Third, we find a bridge
between the north-eastern streamer and the central tidally-unaffected 
part of NGC 4449.  Fourth, during the encounter, material is ejected south 
to NGC 4449. The latter two features can be found in deeper images of 
NGC 4449 (cf.\ Fig.\ 1 of Hunter et al.\ \cite{hunter98}).
However, there are also some remarkable differences:
First, both model and observation show a spur at the lower tip of the 
vertical streamer. However, their directions differ by about 90$^\circ$.
Second, the position of the northern tip of the north-eastern streamer (or
the angle between vertical and the north-eastern streamer) differ.

 Looking at the velocity map, we find Keplerian rotation in the central 
region of the simulations, whereas the observations show rigid rotation. 
The velocities in the streamers reach about 80--100 km\,s$^{-1}$, which
is in agreement with the observed values. Additionally, the velocity
field in the simulations shows a small, but significant V-shape,
south of NGC 4449 which is also observed (cf.\ Fig.\ 3 of Hunter et al.\ 
\cite{hunter98}). The main discrepancies between the observed and 
numerical velocity maps comes from the point mass approximation of the 
galactic potential used for the numerical model. Although this 
approximation must fail for the central region, it should be sufficient
for the outer regions which are most affected by tides 
and which contain almost the entire dynamical mass of NGC 4449.\\

%---------------------------
%  fig4
%---------------------------
\begin{figure}
    \vspace*{8.5cm}
   \caption[ ]{\it Projected particle positions (upper left), 
      velocity map (upper right) and intensity map (lower left) 
      for model B1 ($\epsilon = 0.5$) at the moment of a projected distance
      of 41 kpc. The velocity contours differ by 20 km\,s$^{-1}$, whereas
      the intensity map ranges from 1\% of the maximum intensity 
      $I_{\rm max}$ to $I_{\rm max}$. The star marks the position
      of DDO 125. The observed intensity map
      (cf.\ Fig.\ 1) is displayed in the lower right panel for comparison.}
     \label{reseccdetail}
\end{figure}

\noindent
{\bf Mass ratio.} The variation of the mass ratio gives a lower limit of 
  about 10\% 
  in order to create a sufficient tidal response (Fig.\ \ref{resmass}). 
  If the mass ratio exceeds 0.3, the vertical feature becomes less vertical
  and the north-eastern streamer is more diffuse, both due to the enhanced 
  tidal pull of the intruder.\\

%---------------------------
%  fig5
%---------------------------
\begin{figure}
    \vspace*{12.8cm}
   \caption[ ]{\it Dependence on the mass ratio.
     Projection of the particle positions onto the plane-of-sky at the 
     moment of a projected distance of 41 kpc. The mass ratios are
     - starting with the upper left diagram - 0.01 (model C1), 0.05 (C2),
     0.1 (C3), 0.2 (A, reference model), 0.3 (C4) and 0.4 (C5).
      }
     \label{resmass}
\end{figure}

\noindent
{\bf Minimum distance.} Although the interaction time is not strongly
  influenced by the minimum distance $d_{\rm min}$, it determines the maximum
  strength of the tidal field (Fig.\ \ref{resrmin}):
  For small separations, the structures become more
  diffuse and the primary is much more affected.
  The vertical feature is more 
  extended, creating an additional tidal arm
  (similar to the model with an orbital eccentricity $\epsilon=0.5$).
  If the distance of closest approach exceeds 30 kpc, almost no streamers
  are formed.\\

%---------------------------
%  fig6
%---------------------------
\begin{figure}
    \vspace*{12.8cm}
   \caption[ ]{\it Dependence on the minimum distance during the encounter.
     Projection of the particle positions onto the plane-of-sky at the 
     moment of a projected distance of 41 kpc. The minimum distances are
     - starting with the upper left diagram - 15 kpc (model D1), 20 kpc (D2),
     25 kpc (A, reference model), 30 kpc (D3), 35 kpc (D4) and 40 kpc (D5).
      }
     \label{resrmin}
\end{figure}

\noindent
{\bf Orbital plane.} The orientation of the orbital plane characterizes
  the observer's location or the line-of-sight, which is fixed 
  by two parameters. For the models here only the orbital inclination
  is a free parameter, since the second parameter, the 
  position angle of the orbital plane, is fixed by the
  observed relative line-of-sight velocity of both galaxies.
  The models show that the inclination affects mainly
  the orientation of the vertical streamer (Fig.\ \ref{resiorb}):
  An inclination outside the range $[30^\circ,50^\circ]$ gives too
  much deviation from the south-north direction of the vertical
  feature.
  \\
%---------------------------
%  fig7
%---------------------------
\begin{figure}
    \vspace*{12.8cm}
   \caption[ ]{\it Dependence on the inclination angle of the orbital plane.
     Projection of the particle positions onto the plane-of-sky at the 
     moment of a projected distance of 41 kpc. The inclinations are
     10$^\circ$ (upper left, model E1), 20$^\circ$ (upper right, E2), 
     30$^\circ$ (E3), 40$^\circ$ (A, reference model), 50$^\circ$
     (E4) and 70$^\circ$ (E5).
      }
     \label{resiorb}
\end{figure}

%---------------------------
%  fig8
%---------------------------
\begin{figure}
    \vspace*{12.8cm}
   \caption[ ]{\it 
     Dependence on the extension of the HI distribution in NGC 4449.
     Projection of the particle positions onto the plane-of-sky at the 
     moment of a projected distance of 41 kpc. The radial extensions are
     25 kpc (upper left, model H1), 30 kpc (upper right, H2),
     35 kpc (H3), 40 kpc (A, reference model) and 50 kpc (H4).
      }
     \label{resrhig}
\end{figure}

%---------------------------
%  fig9
%---------------------------
\begin{figure}
    \vspace*{12.8cm}
   \caption[ ]{\it Dependence on the orientation of the HI disc in NGC 4449.
     Projection of the particle positions onto the plane-of-sky at the 
     moment of a projected distance of 41 kpc. The orientation is characterized
     by the inclination $i_d$ of the disc and the position angle $\alpha_d$:
     $i_d = 60^\circ, \alpha_d = 230^\circ$ 
     (upper left, model A, reference model); 60$^\circ$, 210$^\circ$ 
     (upper right, G1); 
     60$^\circ$, 250$^\circ$ (G2); 40$^\circ$, 230$^\circ$ (F1);
     50$^\circ$, 230$^\circ$ (F2) and 70$^\circ$, 230$^\circ$
     (F3).
      }
     \label{resdisc}
\end{figure}

\noindent
{\bf HI distribution in NGC 4449.} A specification of the mass-
and especially the HI distribution is a more difficult task, because, contrary
to the well-known two-body problem, the number of free parameters
characterizing the structure and internal dynamics of the interacting galaxies
is unknown and a matter of investigation itself. For example, it is
not a priori clear how the HI is distributed.
To reduce the number of free parameters, we will
assume here that the observed inner elongated ellipse is the remnant
of an original exponential HI disc, which can be specified by its 
orientation, its scale length and radial extent. 
The inclination of the disc can roughly be determined from the HI map, whereas
its sign is not unambiguous. Placing the south-eastern side of the HI disc into
the foreground means that the orbital plane of the companion
and the plane of the HI disc are nearly orthogonal. We performed several
simulations with geometries of this kind, but these show almost no strong tidal
responses and they all result in situations completely different to that
observed (Kohle \cite{kohle99}). 
However, placing the north-western side of the HI disc into the
foreground and considering the HI velocity field, we now
have the companion on a prograde orbit.

Since the particles are treated as test particles, a variation of the
scale length does not change the final location of a particle, but
rather its mass, which is required for the calculation of the intensity map.
A small scale length (e.g.\ 5 kpc) would confine
too much HI in the center which is unaffected by the encounter. Since
the observations show a large fraction of HI outside 20 kpc,
we have chosen a large scale length of 30 kpc which gives only a
weak radial decline of the HI surface density and, by this, a sufficient
amount of HI in the outer regions.

\noindent
{\sc Disc size.} 
The influence of the maximum radius $R_{\rm max}$ of the disc is 
shown in Fig.\ \ref{resrhig}. If $R_{\rm max}$ is smaller than 35 kpc, the
vertical streamer has not been formed at all, indicating that this feature
stems from the HI gas far outside the center. The north-eastern
arm and the bridge, however, are already built up for $R_{\rm max}=25$ kpc.
With increasing disc size the vertical streamer becomes more
pronounced and more extended, which excludes disc sizes larger than 45 kpc.

\noindent
{\sc Disc orientation.} 
The orientation of the HI disc also strongly affects the appearance
of the primary galaxy. Variation of the position angle $\alpha$ 
by 20$^\circ$ gives HI distributions which are incompatible 
with the observations (Fig.\ \ref{resdisc}). 
If $\alpha$ is decreased, the vertical streamer
is more extended and diffuse. On the other hand, if $\alpha$ is increased,
the north-eastern arm becomes too long and the vertical feature vanishes 
almost completely. 
Similarly, the results depend strongly on the inclination $i$:
Inclinations below $50^\circ$ give only a weak bridge and a weak vertical 
streamer, whereas the orientation of the vertical streamer deviates 
strongly from the
northern direction for inclinations exceeding 70$^\circ$.\\

  The performed parameter study demonstrates that even 
close to the parameters of model A, the final particle 
distribution shows significant variations which are not in
agreement with the observations. Thus, the reference model
might be a good and locally unique candidate describing the 
dynamics of NGC 4449 and DDO 125. 
The confidence in this encounter scenario is especially enhanced
by the strong dependence on the orientation of the HI disc
of NGC 4449. Qualitatively good models are only found if the disc orientation 
agrees well with the values determined independently by observations.
On the other hand, the former parameter
study only investigates a small fraction of parameter space
and, thus, the uniqueness of the solution is not confirmed.
Moreover, in general, it is very tedious to find a good
model by hand, i.e.\ by an unsystematic non-automated search
in parameter space. Thus, the probability of accepting the
first solution found seems to be very high (especially, if
CPU-intensive simulations are performed), by this neglecting
other regions in parameter space which might also give sufficient
fits. In the next section an automatic search in parameter
space by means of a genetic algorithm is introduced. This allows in 
principle for an {\it ab initio} determination of acceptable 
parameter sets (if data sets of sufficient quality are available) or 
at least for a uniqueness test of a favoured scenario like
the reference model described above.

%
%==============================================================
%                 genetic algorithm
%==============================================================
\subsection{The genetic algorithm}
\label{geneticalgorithm}

   The idea of applying models of organic evolution for optimization
problems dates back to the 1960s and 1970s 
(e.g.\ Rechenberg \cite{rechenberg65}; Schwefel \cite{schwefel77}). 
Unlike standard deterministic gradient techniques for optimization (e.g.\ 
the downhill simplex method (Press et al.\ \cite{press92})) 
Rechenberg's {\it Evolutions\-strategie} is probabilistic: 
Starting with a more or less random {\it parent}, i.e.\ a single point in 
parameter space, 
a {\it child} is generated by a random mutation of the parameter set 
characterizing the parent. The quality of both individuals with respect to 
the optimization problem (i.e.\ their {\it fitness\,}) 
determines the parent of the next generation. 
Repeating this process of mutation and selection
improves the quality of the individual monotonously. This
special implementation of evolutionary algorithms
(EAs) in general, has certain advantages: Compared to
complete grids in parameter space, the probabilistic but oriented 
nature of the evolutionary search strategy allows for an efficient check
of a high-dimensional parameter space. Compared to gradient methods
which are very fast near the optimum, EAs do not need any gradient
information which might be computationally expensive. They
depend only weakly on the starting point and - most important - 
they are able to leave local optima. 
The price for these features is a large number of fitness
evaluations or test points in parameter space before converging to a good 
solution.

%---------------------------
%  fig10
%---------------------------
\begin{figure}[h]
    \vspace*{7.0cm}
   \caption[ ]{\it Best fit model during the course of a GA fitting procedure.
       The projection of the particles in $x$-$y$-plane and the corresponding 
       grid for the intensity evaluation are displayed: The original data
       (upper left), the best fit of the GA after initialization (upper middle),
       after the first breeding (upper right), after 11 generations (lower left)
       and at the end of the fitting procedure after 100 generations
       (lower middle). The evolution of the maximum fitness is shown in
       the lower right diagram. The number of test particles per simulation is 900.
           }
   \label{gaevol}
\end{figure}

   Holland (\cite{holland75}) extended the 
{\it Evolutions\-strategie} of Re\-chen\-berg
by abandoning asexual reproduction: He used a population of individuals 
instead of a single individual (and its offspring). From the population
two individuals were selected according to their fitness to become
parents. These parents represent two points ({\it chromosomes}) 
in parameter space
and the corresponding coordinates are treated like genes on a chromosome.
These chromosomes are subject to a {\it cross-over} and a 
{\it mutation} operation
resulting in a new individual which is a member of the next generation.
Such a breeding is repeated until the next generation has been formed. 
Finally, the whole process of sexual reproduction is repeated iteratively
until the individuals representing a set of points in parameter space 
confine sufficiently one or several regions of high fitness. Despite
many differences in the detailed implementations of genetic algorithms (GAs),
the basic concept of an iterated application of randomized 
processes like recombination, mutation, and selection on a population of 
individuals remains in all GAs (Goldberg \cite{goldberg89};
B\"ack \cite{baeck96}). 

%==============================================================
%                  Are the results unique?
%==============================================================
\subsection{Are the results unique?}
\label{unique}

%---------------------------
%  fig11
%---------------------------
\begin{figure}
    \vspace*{11.7cm}
   \caption[ ]{\it Development of six parameters of the best fit model during
     a GA run. The parameters are the mass of the secondary galaxy (upper 
     left), the minimum distance (upper right), the inclination (middle left)
     and position angle (middle right) of the orbital plane, and the
     inclination (lower left) and position angle (lower right) of the 
     disc of the primary galaxy. The filled squares show the parameters
     of the fitted artificial reference model.
      }
     \label{gaparams}
\end{figure}

%---------------------------
%  fig12
%---------------------------
\begin{figure}[t]
    \vspace*{6.2cm}
   \caption[ ]{\it Test of the capability of the GA (lower row) 
    to recover original data
    when these data (upper row) are of low quality or incomplete.
    Shown are runs with low resolution, i.e.\ a discretization on a 
    4$\times$4 grid (left column), a bad pixel (here an omitted central pixel,
    middle column), and a missing structure (the vertical streamer,
    right column).
           }
   \label{gatest}
\end{figure}

  In order to check the uniqueness of the reference model A, several runs 
with different parameter sets encoded in genes were performed. Fig.\ 
\ref{gaevol} shows a typical run of the GA operating on a population of 
100 members. The reference model\footnote{ We were only able to perform a
uniqueness check of our preferred scenario using the best model as reference. A
better model might be determined using the original VLA data as reference.}
 -- displayed in the upper left diagram --
as well as the GA models were calculated with 900 test particles.
This makes the streamers less 
prominent, but still discernible. The next plots show the GA status
at different generations: At generation 1, the best model is
created out of the randomly chosen initial parameter sets. However,
already by the second generation, i.e.\ after the first application of the
reproduction operator, the best model clearly exhibits the north-eastern
streamer. At generation 11, even the weaker vertical feature and the
bridge feature appear. At the end of this GA run the particle distribution
of the original model A and the best fit of the GA are almost identical,
at least compared by eye. The fitness of the best fit has strongly
increased from 0.04 to 0.14 where values above 0.1 indicate 
a very good fit. Most of the improvements were found during the first
50 generations, i.e.\ analyzing 5000 models. After that generation,
the homogeneity of the population led to inbreeding which strongly
prohibits the appearance and spread of new (better) parameters. However,
this is not a problem for the models here, because a very good fit to
all parameters was already found after 40 generations (Fig.\ \ref{gaparams}). 
The relative deviation of the derived
parameters from the original values is less than 15\% in all cases,
and for many is better than 5\%.

  In a series of different GA runs, we varied the population size,
the number of generations or particles, the parameter combination encoded 
on chromosomes, the fitness function, and the random initial population.
In almost all of them the best fit gave an acceptable fit of 
the reference model which demonstrates the capability of GAs
to recover the parameters, even for weak encounters. The models
which failed suffered from inbreeding or from a
fitness function which did not discriminate sufficiently between
high and low quality fits. Moreover, a different region in parameter
space producing a good fit ($f>0.1$) was never found. {\it Thus, 
the model A -- or more exact the corresponding intensity map --
seems to result from a unique parameter set}. However, this
is only a motivated guess, not a mathematical proof!

   Another interesting question is how strongly do the
GA results depend on the completeness or accuracy of the
data. We studied this in two ways: First, we used 
a coarser grid (4x4) in order to mimic a lower resolution of
observational data (left column in Fig.\ \ref{gatest}).
The GA is then only qualitatively able to
reproduce the reference model. Obviously, too much
information is lost when the resolution becomes
too small. On the other hand, an increase of the number of cells to 10x10
did not improve the fit found on a 7x7 grid, but introduced 
more noise into the intensity maps. In the second test
we discarded one or several grid cells from the fitness
calculation (middle and right column in Fig.\ \ref{gatest}).
Omitting one cell does not affect the final fit, 
independent of the location of the cell (here, the center).
However, if a whole feature, such as the vertical streamer,
is neglected, the GA is unable to recover the original
data.

%
%==============================================================
%                  Self-consistent models 'sin gas'
%==============================================================
\subsection{Self-consistent models 'sin gas'}
\label{scnogas}

%---------------------------
%  fig13
%---------------------------
\begin{figure}
    \vspace*{9.5cm}
   \caption[ ]{\it Comparison of restricted N-body models (left) 
     with self-consistent simulations (right) for a parabolic encounter
     (upper row) and an eccentricity of $\epsilon=0.5$ (middle row). The
     projected distance to the perturber (here about 40 kpc) defines the 
     time of the snapshot. The lower row 
     displays the normalized intensity profiles for the self-consistent
     simulations (left: $\epsilon=1$, right: $\epsilon=0.5$). Note that 
     the surface plots are seen from north-west in order to emphasize the 
     coherent streamer structure.     
      }
     \label{resscon}
\end{figure}

%---------------------------
%  fig14
%---------------------------
\begin{figure}
    \vspace*{12.8cm}
   \caption[ ]{\it Self-consistent N-body simulation 'sin gas':
     Projection of the particle positions onto the plane-of-sky at different
     times: initially (upper row), at closest approach (middle row) and 
     at the projected distance of 41 kpc (lower row).
     The left column shows the disc
     particles, whereas the right column displays the halo particles.
      }
     \label{scondiskhalo}
\end{figure}

   The self-consistent models in this paper were performed by a direct
N-body integration using the GRAPE3af board in Kiel for the calculation
of the gravitational forces and the potentials. The time integration was
done by a leap-frog scheme with a fixed timestep of 1.78 Myr. The GRAPE
board uses intrinsically a Plummer softening, the softening length
was chosen to be 0.2 kpc. These choices give a total energy conservation 
of better than 1\% for the whole simulation.

   Different to the restricted N-body models, the set-up of the initial
particle configuration for self-gravitating systems is far from being
trivial. In principle, any stable stationary stellar system fulfills
the collisionless Boltzmann equation. Thus, a solution of the latter
could be used as a trial galactic system. From observations of the 
Milky Way it is known, however, that even if the Galaxy is axisymmetric,
three integrals of motion are required to construct a stationary
model (e.g.\ Binney \& Tremaine \cite{binney87}). Unfortunately, only 
two are known analytically (the energy and the $z$-component of the 
angular momentum). Additionally, even if a particle configuration
is generated from a distribution function depending on the integrals
of motion, it is unclear whether it is stable against
perturbations which might lead to e.g.\ the Toomre instability or 
the bar instability. Thus, on the timescale of interest one has to check 
the stability by numerical simulations.

  In this paper we use the models of Kuijken \& Dubinski (\cite{kuijken95})
consisting of three components which are determined self-consistently
from the Boltzmann equation: the bulge follows a King
model and the halo a lowered Evans model. For the disc the
third integral is approximated by the $z$-energy, i.e.\ the energy 
in vertical oscillations. The advantage of Kuijken \& Dubinski's method
is the possibility to specify parameters like scale length or scale height
of the disc directly, whereas many other methods 
(e.g.\ Barnes \cite{barnes88}) not starting 
in exact virial equilibrium readjust the mass distribution on a
dynamical timescale, which means less control on the structural parameters.

  Our initial model of NGC 4449 is derived from the observational data of 
Bajaja et al.\ (\cite{bajaja94}) and Hunter et al.\ (\cite{hunter98}). 
The total mass is set according to the dynamical mass of 
$7 \cdot 10^{10} \, {\rm M}_\odot$ within 32 kpc. 90\% of this mass is 
attributed to the dark halo, 3\% to the disc and 
7\% to the bulge component. The radial extent of the halo was set to
32 kpc. The disc modelled here should not be confused
with the optical disc of the galaxy. Since we are interested in the
tidal response of the outer regions, especially the formation of the
streamers, we did not model the inner (optical) region of NGC 4449,
i.e.\ the stellar disc. The missing signatures of any tidal features in 
the optical image demonstrate that the stellar body of NGC 4449 is
not affected by the interaction. For the same reason we did not consider
the counter-rotation of the inner region. However, its gravity is taken
into account by means of the bulge and halo contributions to the
overall rotation curve. The radial extension of the bulge is 3 kpc.
The scale length of the extended (HI) disc was varied
between 10 and 30 kpc, the outer edge was set to 40 kpc.

The long-term stability was checked by simulations of an isolated
disc with $N=18000$ particles (8000 for the disc, 4000 for the bulge 
and 6000 for the halo). Within 1.5 Gyr the Lagrange radii at 10\% and
90\% mass vary only by 2-4\%. They show no systematic trend except for a
small expansion during the first 100 Myr which is probably caused by
gravitational softening. Due to two-body relaxation the scale height of
the disc increases by almost a factor of 2 within 1.5 Gyr. However, this
heating is not expected to alter the result of the dominant interaction 
significantly. The Fourier amplitude $C_2$, defined by 
$C_m \equiv (\int_{\rm disc} \Sigma(R,\phi) r \, dr \, e^{-im\phi} d\phi) /
            (\int_{\rm disc} \Sigma(R,\phi) r \, dr \, d\phi))$ (with the
surface density $\Sigma$), is always less than 5\%, in agreement
with no significant bar or spiral pattern built up during the simulation.

   Fig.\ \ref{resscon} shows a comparison of restricted N-body 
simulations with self-consistent models. Except for
the more diffuse structure in the
latter, the large-scale features are very similar in both simulations.
The streamer structure can be found in both self-consistent 
simulations, as the surface plots demonstrate. The fact that the streamers
are better resolved in the simulation with an eccentricity 
$\epsilon = 0.5$, is partly due to the increased number of particles
(80\,000) used for this simulation. The simulations for the parabolic orbits 
were performed with 8\,000 (disk) particles for both, the restricted
and the self-consistent simulations. The latter clearly demonstrates
the ability of the restricted N-body models to resolve structures, even
when a small number of particles is used. It is remarkable that the 
extended dark halo which has been applied for the self-consistent models
does not strongly affect the results as the comparison
of the $\epsilon = 0.5$-models shows: The vertical and the north-eastern 
streamer, the mass ejected to north of the north-eastern streamer,
the small bridge between the central region and the eastern tip of the
streamer, the extent of the HI distribution south and south-east to
the center show up in both simulations at almost the same locations.
The only significant difference is that the north-eastern streamer is less
straight in the self-consistent model than in the restricted N-body
simulation. This very good agreement confirms the applicability of the
restricted N-body method here. Another interesting aspect 
concerns the response of the halo
to the intruder. Unlike to the disc, the halo remains almost 
unchanged by the encounter (Fig.\ \ref{scondiskhalo}). This
behaviour allows the application of a rigid galactic potential
which has been used in the restricted N-body simulations.

%
%==============================================================
%                  Self-consistent models 'con gas'
%==============================================================
\subsection{Self-consistent models 'con gas'}
\label{stickyparticles}

%---------------------------
%  fig15
%---------------------------
\begin{figure}
    \vspace*{12.8cm}
   \caption[ ]{\it
     Projection of the particle positions onto the plane-of-sky at different
     times: at closest approach (upper row), at the projected distance of
     41 kpc (middle row) and at the end of the simulation (lower row).
     Both models include self-gravity. The left column shows a purely
     stellar simulation, whereas the right column includes gas dynamics
     by means of sticky particles.
      }
     \label{scondissnodiss}
\end{figure}

    For the gasdynamical models we use a sticky particle scheme which 
treats each particle of the HI disc as a gaseous cloud. The radius of
these clouds is assumed to follow $R_{\rm cl} = 20 \cdot 
\sqrt{M_{\rm cl} / 10^6 \, {\rm M}_\odot} \, {\rm pc}$ which is derived from
the mass-radius relation of Rivolo \& Solomon (\cite{rivolo87}) corrected
for glancing collisions ($M_{\rm cl}$ is the mass of a gas cloud). 
When the distance of two clouds falls below
the sum of their radii, merging
of the clouds is allowed, if their relative orbital angular momentum
is smaller than the maximum angular momentum a merged cloud
can acquire before break-up. If this angular momentum criterion prevents 
a merger, the clouds keep their kinematical data, i.e.\ no energy is 
dissipated. Star formation is only qualitatively included by adopting
an upper mass limit of $7 \cdot 10^{8} \, {\rm M}_\odot$ for clouds. If they
exceed this limit no further merging is allowed, i.e.\ they are treated
as stars. However, the star formation criterion is not important
for the models of this paper. (For details of the cloud model or the
merging mechanism see Theis \& Hensler \cite{theis93}). This dissipative
scheme has been successfully applied to collapsing systems reproducing
many observed properties of spiral and elliptical galaxies. For the
long-term evolution of barred disc galaxies, a comparison of this scheme with 
that of Palou\v s et al.\ (\cite{palous93}) for their standard coefficients
of restitution of $\beta_r = \beta_t = 0.2$ gives 
basically the same results (Jungwiert, private communication). 
Thus, the chosen model
should be applicable for a reliable determination of the difference
between gas dynamical and purely stellar dynamical models.

     A comparison of gaseous models with dissipationless models
shows no significant difference (Fig.\ \ref{scondissnodiss}). Due
to the low gas densities only about 20-30\% of the clouds 
undergo inelastic collisions which mainly occur in the
denser central regions. Thus, the outer HI halo is only very
weakly influenced by (sticky particle) gas dynamics and the applicability
of purely stellar simulations for the HI is possible.
%It would be interesting to compare these models with an SPH
%approach emphasizing the diffuse nature of the interstellar medium.

%
%==============================================================
%                  Discussion
%==============================================================
\section{Discussion}
\label{discussion}

\subsection{Interaction between NGC 4449 and DDO 125?}

   The previous simulations demonstrate that models based on an encounter 
scenario are able to reproduce the morphology of the three streamers.
Furthermore, the interaction puts gas to the south-west of 
NGC 4449, where a clumpy HI distribution is observed (Hunter et al.\
\cite{hunter98}).
Though its detailed patchy structure has not been recovered
by the models, an initial gas distribution less regular than the 
smooth disc-like one used for the numerical models might 
be responsible for this deviation. 

   The parameter study showed that small variations
in the parameters give HI distributions which are not in agreement
with the observations. Especially, the strong dependence of the
final structure on the initial orientation of the HI disc
confines its inclination and position angle within 10$^\circ$
to the values derived {\it independently} from the observations. 
This remarkable coincidence would occur necessarily in the
encounter scenario, whereas in the infall scenario it happens
just by chance. A second critical feature is the abrupt direction 
change of the streamers. In the infall scenario they mark
the path of the perturbed gas which 
collapses to the galactic center. However, it is hard to
imagine how a realistic static potential could create orbits 
with such strong
direction changes as e.g.\ the one from the vertical to the bridge
feature. Moreover, due to the large distance of these
features from the galactic center the potential should
be more or less spheroidal or already spherical, which makes
it even more difficult to produce such orbits.

  The most critical parameter for the interaction scenario is the
mass ratio $q$ of DDO 125 and NGC 4449. The simulations demonstrate that
a mass ratio smaller than 10\% cannot reproduce the streamer structure.
In order to determine $q$, both galactic masses have to be
known. In Bajaja et al.'s (\cite{bajaja94}) estimate for the mass of NGC 4449 
they assumed that the HI is in rotational equilibrium even at a
distance of 32 kpc. However, in the case of an interaction with DDO 125
the velocities at large distances are strongly affected by
the interaction and they are definitely not in rotational equilibrium
(as the ejection of the tidal arms in a continuation of the simulations 
demonstrates). Hence, the derived mass of $7 \cdot 10^{10} \, {\rm M}_\odot$ is 
at best an upper mass limit of NGC 4449. A better guess
is the dynamical mass inside the central gaseous ellipse
which seems to be unaffected by interaction. E.g.\ the dynamical
mass inside 11 kpc would be a factor 3 smaller than Bajaja's value.

  Even more difficult is the mass determination
of DDO 125. Observations by Tully et al.\ (\cite{tully78}) gave a dynamical
mass of $5 \cdot 10^8 \, {\rm M}_\odot$ inside the optical radius of DDO 125.
However, Ebneter et al.\ (\cite{ebneter87}) reported a linear increase
in the rotation curve exceeding the range observed by Tully et al.
by a factor of two. Moreover, they found no hint of a flattening
or turn-over in DDO 125's rotation curve. This gives a lower mass 
limit of $4 \cdot 10^9 \, {\rm M}_\odot$. Combining the mass estimates yields
a lower limit of the mass ratio of 6\% when using Bajaja's mass 
determination of NGC 4449. A more realistic mass ratio might be
derived from the ranges of rigid rotation resulting
in $q\approx18\%$. Thus, observationally it cannot be excluded that
the mass ratio is below the critical one for the interaction
scenario, however, the more realistic estimates seem to be in 
agreement with the values found in the simulations. The situation
could be clarified when a reliable mass determination of DDO 125 
is available.

\subsection{Origin of the extended HI-distribution around NGC 4449}

   Another important question is related to the initial conditions
of the previous simulations, i.e.\ the origin of the extended
HI distribution. In principle, there are several possibilities:

\noindent
{\it Idea 1. The HI comes from DDO 125.}
In a previous encounter ($\epsilon<1$) the gas was stripped off 
from DDO 125. Due to the long orbital period of e.g.\ 3.6 Gyr for 
$\epsilon=0.5$ the gas has enough time to settle down in a regular 
disc-like structure. However, the amount of HI gas in DDO 125 is much
smaller than the gas seen around NGC 4449. It seems unlikely that
DDO 125 could have survived the loss of such a large fraction of its 
gas content without disruption. 

\noindent
{\it Idea 2. The HI comes from NGC 4449.}
Though a star forming dwarf galaxy drives a galactic 
wind e.g.\ by means of type II supernovae, a detailed fine-tuning 
of the energy injection would be required to prevent 
a blow-out of the HI gas, especially if it acquired enough energy
to form such an extended system. Moreover, the 
counter-rotation of the HI outside and inside the optical
radius of NGC 4449 makes this scenario not very appealing.

\noindent
{\it Idea 3. The HI gas stems from a previous minor merger with
NGC 4449}. Though a satellite galaxy can be
disrupted by the tidal field of NGC 4449, it is very unlikely
that it is already disrupted far outside the galactic
center where the HI gas is detected.

\noindent
{\it Idea 4. The HI originates from the collapse of a dwarf companion.}
A dwarf galaxy forms in the vicinity of NGC 4449 similar to
the formation scenario of compact ellipticals suggested by
Burkert (\cite{burkert94}).
By this it undergoes a strong collapse which leads to mass ejection 
due to violent relaxation. In principle, a mass loss of up to 40\% of
the companion could occur, providing enough gas to explain the HI.
However, where is the dwarf today? Excluding an
unlikely radial orbit, the dwarf's orbit could in principle 
decay by dynamical friction. However, due to the large distance
to NGC 4449 the decay time should be at least several orbital periods,
which exceeds the Hubble time.

\noindent
{\it Idea 5. Another major encounter.} Although NGC 4449 is a member
of the loose Canes Venatici group, there seems to be
no close galaxy which has a radial velocity in agreement with
the rotational sense of the gaseous disc in NGC 4449. The only candidate is 
the ringed Sab-type galaxy NGC 4736, located about 5.5 degrees south-east of 
NGC 4449. The projected distance of 350 kpc and the differential radial 
velocity of \mbox{$\Delta v=101$ km s$^{-1}$} (NED) is compatible with a 
scenario in which both galaxies experienced a close encounter about 
$3.5 \cdot 10^9$ years ago on a hyperbolic orbit (Kohle \cite{kohle99}). 
Placing NGC 4736 at a distance of 4.4 Mpc, its total HI mass is
$4.2 \cdot 10^8 \bf{M_\odot}$ (Bosma et al.\ \cite{bosma77}; 
Mulder \& van Driel \cite{mulder93}), which is quite low for Sab-type galaxies 
(Solanes et al.\ \cite{solanes96}). Might it be that NGC 4449 formed its 
halo out of gas stripped from another galaxy?

%
%==============================================================
%                  Conclusions
%==============================================================
\section{Summary and conclusions}
\label{conclusions}

 Several N-body methods are combined in order to develop
a method for the determination of the parameters of interacting
galaxies. This method is applied to the HI distribution
of NGC 4449. 
In a first step, the fast restricted N-body method is used
to confine a region in parameter space which reproduces the main
observed features. In a second step, a genetic algorithm (also
using restricted N-body calculations) is employed
which allows, in principle for both, an automatic fit of observational
data even in a high-dimensional parameter space and/or 
a uniqueness test of a favoured parameter combination (only the latter
has been done in this paper). For a
genetic algorithm one typically has to follow a population of (at least) 100
members for 100 generations in order to get a good fit, provided the
data are sufficiently accurate. Missing single pixels do not
inhibit the parameter determination, as long as the key features are included
in the data. Since a typical restricted N-body simulation takes
a few CPU-seconds on a Sparc workstation, the whole fitting
procedure is finished after 3--6 CPU-hours.

  In the third step, the results of the previous steps
are compared with detailed self-consistent N-body simulations.
In the case of NGC 4449, they show that the restricted
N-body calculations are reliable models for this encounter. 
The comparison with the sticky particle models demonstrates
that the HI gas can be modeled without any restriction
by a purely stellar dynamical approach, provided the encounter
is weak, as in the case of NGC 4449 and DDO 125.

 From the previous simulations, we conclude that the extended
HI features observed in NGC 4449 are created by an encounter
with DDO 125. Prior to the encounter, the HI gas formed
an extended, almost homogenous disc with a large scale length 
of about 30 kpc and a radial extension of 40 ($\pm10$) kpc.
The orientation of the disc is in agreement with the
orientation of the inner ellipsoidal HI distribution.
The orbital plane has an inclination
angle of about 40$^\circ$ ($\pm 10^\circ$). The eccentricity
is in the vicinity of a parabolic encounter, but favouring the region
of elliptical orbits ($0.5<\epsilon<1.5$).
The apocenter distance of the galaxies is about
25 ($\pm5$) kpc. The closest approach happened $3.5-6.2 \cdot 10^8$ yr ago
(depending on eccentricity). The mass ratio of both galaxies must
be approximately 0.2 (and cannot be smaller than 0.1).
 
\begin{acknowledgements}
 The simulations were partly performed with the
 GRAPE3af special purpose computer in Kiel (DFG Sp345/5).
 The authors are grateful to Deidre Hunter, Jay Gallagher, Uli Klein
 and Hugo van Woerden for stimulating discussions about NGC 4449.
 We are also greatly indebted to Konrad Kuijken and John Dubinski who made 
 their program for the generation of initial particle distributions available
 to the public, and Paul Charbonneau and Barry Knapp for providing their 
 {\sc pikaia} code. S.K. acknowledges DFG grant III GK-GRK 118/2. 
 This research has made use of the NASA/IPAC Extragalactic Database (NED) 
 which is operated by the Jet Propulsion Laboratory, California Institute of 
 Technology, under contract with the National Aeronautics and Space 
 Administration. Finally, we thank the language editor of A\&A for a careful
 reading (and improving) of the manuscript. 
\end{acknowledgements}

%#######################################################
%               BIBILIOGRAPHY
%#######################################################

%========================
%    REFERENCES
%========================

\appendix
%
%==============================================================
%                 genetic algorithm
%==============================================================
\section{The genetic algorithm}
\label{appendixgeneticalgorithm}

In the following sections we briefly describe the main ingredients
of our genetic algorithm. The GA used here is a slightly 
modified version of P.\ Charbonneau's code {\sc pikaia} (for details see 
Charbonneau (\cite{charbonneau95})). The interface between the GA
and the simulations, i.e.\ the fitness determinations, is performed in 
a similar way as in Wahde (\cite{wahde98}), however the coding of 
genes as well as the choice of parameters used as genes differ from
Wahde's approach.

%---------------------------
%  fig16
%---------------------------
\begin{figure}[t]
   \psfig{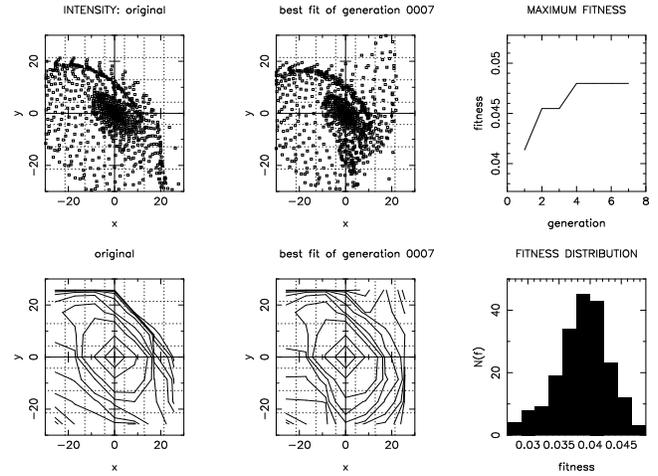}
   \caption[ ]{\it Fitness and best fit of an early generation during the GA
       fitting procedure. The figure displays the artificial original data
       (left column: positions of the particles in the $x$-$y$-plane
        (upper left) and the resulting intensity contour lines
        (linear spacing, lower left), the best fit of the actual generation
        (middle column) and information about the fitness history, i.e.\
        the maximum fitness (upper right) and the actual fitness distribution
        (lower right). The underlying grid in the left and middle column
        is used to determine the intensities.
           }
   \label{fig3}
\end{figure}

\subsection{Fitness and parent selection}

 In order to determine the fitness of an individual (i.e.\ a special parameter
set), the N-body simulation has to be performed and compared with the
observation. Generally, this is done by mapping the particles on a cartesian 
grid. However, here a quadratic grid of size 60 kpc centered on NGC 4449
is used. The resolution
of the grid was set to 7x7 in order to allow for a better resolution of the
tidal features of an encounter. For each grid cell the intensity was
calculated by summing the masses of the individual particles in that
cell. The conversion to observed intensities can be performed by assuming
a mass-to-light ratio for the particles (or treating it as another free 
parameter) or by normalizing the intensities. Here the intensities are
normalized to the total intensity of the whole grid. The quality can be
measured by the relative deviation $\delta$ of the intensities in both 
maps:
\begin{equation}
                \delta \equiv 
                     \sum_{i ({\rm cells})} \, \, \, 
                     \frac{|I_{{\rm ref},i} - I_{{\rm mod},i}|}
                      {\max(I_{{\rm ref},i},I_{{\rm mod},i})}
\end{equation}

The fitness $f$ is then estimated by
\begin{equation}
         f \equiv \frac{1}{1+\delta}\,. \hfill
\end{equation}

\noindent
  In the case of an impossible parameter configuration (e.g.\ a circular orbit
with a distance less than the observed projected distance) or any convergence
problems the fitness of that point in parameter space is set to zero, 
prohibiting any offspring.
Typically, only a few models in $10^4$ parameter sets fail,
most of them in the first few generations before the system starts to 
approach the parameter space region of interest.

  In order to determine the parents of the next generation, the members
of the actual generation are ranked by their fitness. The parents are then
selected randomly, whereas the probability of being chosen is proportional
to the rank ({\it roulette-wheel selection}).

  A typical example of the status of the GA after a few generations
is shown in Fig.\ \ref{fig3}. In the left column the positions of
the original data are discretized on the grid, yielding the intensities
displayed on the lower left. Though the resolution of the
grid (about 8.6 kpc) exceeds the thickness of the tidal 
structures, the large scale pattern is quickly found:
in the middle column the best fit of the actual (here 7th)
generation already agrees qualitatively with the original. 
This demonstrates that large scale features already present in intermediate
resolution maps can be sufficient to constrain the dynamics. This is in
agreement with Wahde's (1998) calculations for equal mass galaxies:
he recovered orbital parameters by using a 4$\times$4 grid. 
The quality of our fits is calculated only by a comparison of the lower 
two intensity 
diagrams. The fitness distribution in the population resembles a Gaussian
at the beginning of the fitting procedure.

\subsection{Coding of parameters}
%---------------------------
%  fig17
%---------------------------
\begin{center}
\begin{figure}[ptbh]
   \psfig{figure=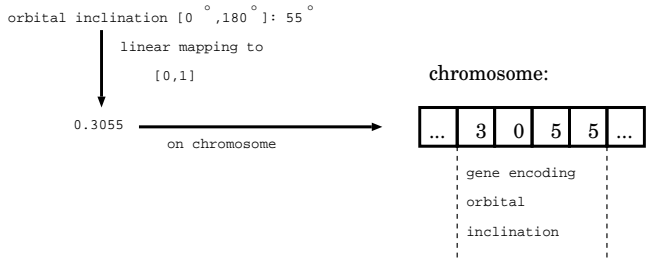,angle=270,height=3.5cm,width=8.5cm}
   \caption[ ]{\it Schematic diagram for the coding of parameters on
      a numerical chromosome.
           }
   \label{coding}
\end{figure}
\end{center}

  In organic systems the information is coded in genes in the form of
sequences of the four nucleotide bases where a triplet of them (a codon) 
encodes one amino acid, the building block of proteins. 
The genes themselves are 
arranged in one or several chromosomes which contain all the genetic
information (or the genotype). The phenotype which is subject to the
selection process is the result of the genetic information and the
interaction with its environment. 

   Translated to our fitting problem, the
phenotype is the final intensity map created by a special choice of
parameters or initial conditions. The latter correspond to the proteins
which physically express the genetic information. The mapping between
the parameters and the genes or the number of chromosomes is not fixed in GAs. 
Many GAs apply a binary alphabet to encode the parameters
(e.g.\ Holland \cite{holland75}; Goldberg \cite{goldberg89}). 
In Charbonneau's program
{\sc pikaia} a  decimal encoding is implemented. For our calculations we
combined four decimal digits giving a number between zero and one.
This {\it gene} has been mapped to a real parameter by a linear or logarithmic 
transformation, e.g.\ the orbital inclination is mapped
linearily to the range [$0^\circ$, 180$^\circ$] (Fig.\ \ref{coding}). 
In general, the linear mapping of parameter $x$, which is allowed to vary 
between $x_{\rm low}$ and $x_{\rm high}$, is performed by
\begin{equation}
         x = x_{\rm low} + g \cdot (x_{\rm high} - x_{\rm low})
\end{equation}
Here $g$ denotes the value of the gene which is in the range
[0,1]. Similarly, the logarithmic mapping is done by replacing
$x$ by $\log(x)$.
All the genes are assumed to be located in a linear fashion on a single 
chromosome. Unlike organic evolution, the phenotype is completely 
determined by the genotype, no environmental influence on the individual 
acts after its genotype is fixed.

\subsection{Cross-Over}

%---------------------------
%  fig18
%---------------------------
\begin{center}
\begin{figure}[ptbh]
   \psfig{figure=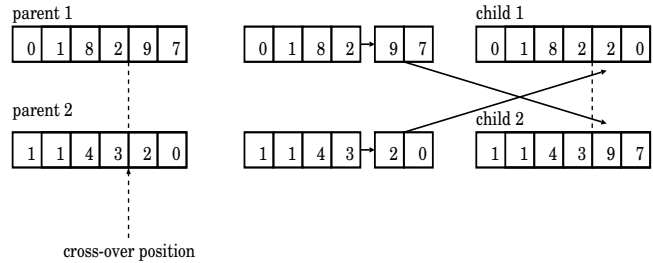,angle=270,height=3.5cm,width=8.5cm}
   \caption[ ]{\it Schematic diagram for the cross-over operation.}
   \label{crossover}
\end{figure}
\end{center}

   The main difference between evolutionary strategies and
genetic algorithms is the sexual reproduction in the framework of GAs. 
This means 
that the chromosomes of the parents are combined in a new manner by means of
a cross-over mechanism in addition to mutation. The principle idea 
is illustrated in Fig.\ \ref{crossover}: After determining the two
parents a random position on the chromosomes is selected. At this position
the chromosomes are split into two fragments which are exchanged
among the chromosomes resulting in two new ones.

\subsection{Mutation}
%---------------------------
%  fig19
%---------------------------
\begin{center}
\begin{figure}[ptbh]
   \psfig{figure=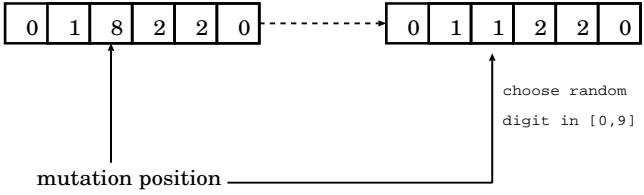,angle=270,height=2.5cm,width=8.5cm}
   \caption[ ]{\it Schematic diagram for the mutation process.}
   \label{mutation}
\end{figure}
\end{center}

   In order to introduce completely new information into a genepool
a mutation process, i.e.\ a random change of the information on the
chromosomes is introduced (Fig.\ \ref{mutation}). For each position
on a chromosome, mutation is applied with a small probability
(typ.\ $p_{\rm mut} \approx 0.5\%$). After selecting a position for
mutation, an integer random number in the range [0,9] replaces
the original value at that position. Thus, a real change in the genetic
information takes place with a probability of 90\%, even if
that position of a gene is subject to mutation. The overall probability
for at least one real mutation on a chromosome coding 6
parameters by 4 digits is about 11\%. Thus, cross-over is the most
important process introducing diversity into the population, whereas
mutation affects only each 9th member of the population. In principle,
the mutation rate can be increased, but the price would be a more
or less random search in parameter space which is already after a 
few generations inferior to the oriented search of a GA method
(Wahde \cite{wahde98}).

   A principal danger of GAs is inbreeding, i.e.\
the homogenization of the genepool by one or a few dominant 
genes. If they successfully replace other genes, cross-over does not
introduce sufficient (or -- if inbreeding is complete -- any) diversity 
into the population and thus any further improvement of the
chromosomes is prohibited. The only way to overcome this situation
is to introduce more diversity by mutation. Therefore, {\sc pikaia}
allows for a variable mutation rate which increases the mutation
probability $p_{\rm mut}$ whenever the fitness distribution becomes narrow,
i.e.\ inbreeding is suspected to operate.
On the other hand, $p_{\rm mut}$ is decreased if the fitness distribution
becomes very broad, i.e.\ the randomization is dominant.

\subsection{Determination of the next generation}

   After $N_{\rm pop}$ children have been created, the old
generation is completely replaced by the next generation ({\it full 
generational replacement}). The only exception is the survival of the fittest
member of the parent generation, if it is fitter than the child which 
would replace it ({\it elitism}). This mechanism prevents the
system from forgetting successful solutions. 

  The initial population is randomly chosen in the accessible
parameter space. A standard population size is about 
$N_{\rm pop} = 100 - 200$ members. The GA converges typically after
about $N_{\rm gen} = 100$ generations. Thus, $10^4$ or a few $10^4$
N-body simulations are required for a single GA fit. For a
genetic algorithm simulation with 900 particles (and no offset for an inner
edge) 2 CPU-seconds are required per simulation compared to 3 CPU-hours 
on a GRAPE3. Thus, self-consistent
N-body models, such as the mentioned GRAPE models, would still need
3.4 years, whereas the restricted N-body models reduce the
CPU-requirement to 5.6 hours on a 150MHz-Sparc10.


\begin{thebibliography}{}

\bibitem[1966]{arp66} 
  Arp H., 1966, {\it Atlas of peculiar galaxies}, Pasadena, Caltech
          (also Arp H., 1966, ApJS, 14,1)

\bibitem[1996]{baeck96} 
  B\"ack Th., 1996, {\it Evolutionary Algorithms in Theory and Practice},
          Oxford Univ.\ Press, Oxford

\bibitem[1994]{bajaja94}  
  Bajaja E., Huchtmeier W.K., Klein U., 1994, A\&A, 285, 388

\bibitem[1988]{barnes88}  
  Barnes J., 1988, ApJ, 331, 699

\bibitem[1996]{barnes96}  
  Barnes J., Hernquist L., 1996, ApJ, 471, 115

\bibitem[1986]{barnes86}  
  Barnes J., Hut P., 1986, Nature, 324, 446

\bibitem[1987]{binney87}
  Binney J., Tremaine S., 1987, {\it Galactic Dynamics}, Princeton
   Univ.\ Press, Princeton

\bibitem[1997]{bomans97}
  Bomans D., Chu Y.-H., 1997, AJ, 113, 1678

\bibitem[1977]{bosma77}
  Bosma A., van der Hulst J.M., Sullivan W.T., 1977, A\&A, 57, 373

\bibitem[1994]{burkert94}
  Burkert A., 1994, MNRAS, 266, 877

\bibitem[1982]{casoli82}
  Casoli F., Combes F., 1982, A\&A, 110, 287

\bibitem[1997]{dellaceca97}
  della Ceca R., Griffiths R.E., Heckman T.M., 1997, ApJ, 485, 581

\bibitem[1995]{charbonneau95}
  Charbonneau P., 1995, ApJS, 101, 309

\bibitem[1969]{crillon69}
  Crillon R., Monnet G., 1969, A\&A, 1, 449

\bibitem[1987]{ebneter87}
  Ebneter K., Davis M., Jeske N., Stevens M., 1987, BAAS, 19, 681

\bibitem[1981]{fisher81}
  Fisher J.R., Tully R.B., 1981, ApJS, 47, 139

\bibitem[1989]{goldberg89}
  Goldberg D.E., 1989, {\it Genetic Algorithms in Search, Optimization,
         \& Machine Learning}, Addison-Wesley, Reading

\bibitem[1987]{hernquist87}
  Hernquist L., 1987, ApJS, 64, 715

\bibitem[1990]{hernquist90}
  Hernquist L., 1990, J. Comp.\ Phys., 87, 359

\bibitem[1989]{hernquist89}
  Hernquist L., Katz N., 1989, ApJS, 70, 419

\bibitem[1992]{hernquist92}
  Hernquist L., Ostriker J.P., 1992, ApJ, 386, 375

\bibitem[1994]{hill94}
  Hill R.S., Home A.T., Smith A.M., Bruhweiler F.C., Cheng K.-P., 
          Hintzen P.M., Oliversen R.J., 1994, ApJ, 430, 568

\bibitem[1975]{holland75}
  Holland J., 1975, {\it Adaptation in natural and artificial systems},
         Univ.\ of Michigan Press, Ann Arbor

\bibitem[1941]{holmberg41}
  Holmberg E., 1941, ApJ, 94, 385

\bibitem[1997]{hunter97}
  Hunter D.A., Gallagher J.S., 1997, ApJ, 475, 65

\bibitem[1986]{hunter86}
  Hunter D.A., Gillet F.C., Gallagher J.S., Rice W.L., Low F.J., 1986, 
         ApJ, 303, 171

\bibitem[1996]{hunter96}
  Hunter D.A., Thronson H.A., 1996, ApJ, 461, 202

\bibitem[1998]{hunter98}
  Hunter D.A., Wilcots E.M., van Woerden H., Gallagher J.S.,
          Kohle S., 1998, ApJL, 495, L47

\bibitem[1755]{kant55}
  Kant I., 1755, in {\it Allgemeine Naturgeschichte und Theorie des
         Himmels oder Versuch von der Verfassung und dem mechanischen
         Ursprunge des ganzen Weltgeb\"audes nach Newtonischen
         Grunds\"atzen abgehandelt}, ed.\ H. Ebert, 1890, 
         Verlag Wilhelm Engelmann, Leipzig

\bibitem[1998]{karachentsev98}
  Karachentsev I.D., Drozdovsky I.O., 1998, A\&AS, 131, 1

\bibitem[1996]{klein96}
  Klein U., Hummel E., Bomans D., Hopp U., 1996, A\&A, 313, 396

\bibitem[1998]{kohle98}
  Kohle S., Klein U., Henkel C., Hunter D.A., 1998, in
        Proc.\ of {\it The Magellanic Clouds and Other Dwarf Galaxies},
         ed.\ T. Richtler \&  J.M. Braun, Bad Honnef, p.\ 265

\bibitem[1999]{kohle99}
   Kohle S., Ph. D. thesis, University of Bonn, 1999

\bibitem[1995]{kuijken95}
  Kuijken K., Dubinski J., 1995, MNRAS, 277, 1341

\bibitem[1993]{mulder93}
  Mulder P., van Driel W., 1993 A\&A, 272, 63 

\bibitem[1988]{noguchi88}
  Noguchi M., 1988, A\&A, 203, 259

\bibitem[1993]{palous93}
  Palou\v s J., Jungwiert B., Kopeck\'y J., 1993, A\&A, 274, 189

\bibitem[1961]{pfleiderer61}
  Pfleiderer J., Siedentopf H., 1961, Zs.\ f.\ Ap., 51, 201

\bibitem[1992]{press92}
  Press W.H., Teukolsky S.A., Vetterling W.T., Flannery B.P., 1992,
        {\it Numerical Recipes in FORTRAN - The Art of Scientific Computing},
         Cambridge Univ.\ Press, Cambridge

\bibitem[1965]{rechenberg65}
  Rechenberg I., 1965, {\it Cybernetic solution path of an experimental
         problem}, Royal Aircraft Establishment, Library Translation No.\ 1122,
         Farnborough

\bibitem[1987]{rivolo87}
  Rivolo A.V., Solomon P.M., 1987, in {\it Molecular Clouds in the 
     Milky Way and External Galaxies}, ed.\ R. L. Dickman et al., 
     Springer, p.\ 42

\bibitem[1990]{sasaki90}
  Sasaki M., Ohta K., Saito M., 1990, PASJ, 42, 361

\bibitem[1977]{schwefel77}
  Schwefel H.-P., 1977, {\it Numerische Optimierung von 
     Computer-Modellen mittels der Evolutionsstrategie}, Birkh\"auser, Basel

\bibitem[1980]{sellwood80}
  Sellwood J.A., 1980, A\&A, 89, 296

\bibitem[1987]{silk87}
  Silk J., Wyse R.F.G., Shields G.A., 1987, ApJ, 322, L59

\bibitem[1996]{solanes96}
  Solanes J.M., Giovanelli R., Haynes M.P., ApJ, 461, 609

\bibitem[1990]{sugimoto90}
  Sugimoto D., Chikada Y., Makino J., Ito T., Ebisuzaki T.,
          Umemura M., 1990, Nature, 345, 33

\bibitem[1993]{theis93}
  Theis Ch., Hensler G., 1993, A\&A, 280, 85

\bibitem[1998]{theis98}
  Theis Ch., Kohle S., 1998, in
        Proc.\ of {\it The Magellanic Clouds and Other Dwarf Galaxies},
         ed.\ T. Richtler \&  J.M. Braun, Bad Honnef, p.\ 209 

\bibitem[1993]{thomasson93}
  Thomasson M., Donner K.J., 1993, A\&A, 272, 153

\bibitem[1972]{toomre72}
  Toomre A., Toomre J., 1972, ApJ, 178, 623 

\bibitem[1978]{tully78}
  Tully R.B., Bottinelli L., Fisher J.R., Gouguenheim L., Sancisi R.,
          van Woerden H., 1978, A\&A, 63, 37

\bibitem[1975]{vanwoerden75}
  van Woerden H., Bosma A., Mebold U., 1975, in {\it La Dynamique
          des Galaxies Spirales}, ed.\ L. Weliachew, p.\ 483

\bibitem[1850]{humboldt50}
  von Humboldt A., 1850, in {\it Kosmos}, Vol.\ III, Cotta'scher Verlag,
  Stuttgart, p.\ 178

\bibitem[1959]{vorontsov59}
  Vorontsov-Velyaminov B.A., 1959, {\it Atlas and Catalog of
          Interacting Galaxies}, Moscow, Sternberg Institute

\bibitem[1998]{wahde98}
  Wahde M., 1998, A\&AS, 132, 417 

\end{thebibliography}
\end{document}